\documentclass[%
12pt,
aip,
cha,
amsmath,amsfonts,amssymb,
author-numerical,
reprint,
floatfix,
]{revtex4-2}

\usepackage[utf8]{inputenc}
\usepackage[T1]{fontenc}
\usepackage{graphicx}
\usepackage{hyperref}
\usepackage{mathrsfs}

\renewcommand{\d}[1]{\ensuremath{\operatorname{d}\!{#1}}}

\def\const{\mathrm{const}}

\def\u{\mathbf u}
\def\k{\mathbf k}

\def\p{\varphi}

\def\PS{\Phi^\mathrm{S}}
\def\PU{\Phi^\mathrm{U}}

\def\H{\mathcal H}
\def\E{\mathcal E}
\def\F{\mathcal F}
\def\G{\mathcal G}
\def\C{\mathcal C}
\def\M{\mathcal M}

\def\Fp{\frac{\delta\mathcal F}{\delta\p}}

\def\Fb{\frac{\delta\mathcal F}{\delta\bar\xi}}

\def\Fs{\frac{\delta\mathcal F}{\delta\psi_\sigma}}

\def\Gb{\frac{\delta\mathcal G}{\delta\bar\xi}}

\def\Gs{\frac{\delta\mathcal G}{\delta\psi_\sigma}}

\begin{document}

\title{Nonlinear saturation of thermal instabilities}

\author{F.\ J.\ Beron-Vera} \email{fberon@miami.edu}
\affiliation{Department of Atmospheric Sciences, Rosenstiel School
of Marine and Atmospheric Science, University of Miami, Miami,
Florida 33149, USA}

\date{\today}
\begin{abstract}
  Low-frequency simulations of a one-layer model with lateral
  buoyancy variations (i.e., thermodynamically active) have revealed
  circulatory motions resembling quite closely submesoscale
  observations in the surface ocean rather than indefinitely growing
  in the absence of a high-wavenumber instability cutoff. In this
  note it is shown that the existence of a convex pseudoenergy--momentum
  integral of motion for the inviscid, unforced dynamics provides
  a mechanism for the nonlinear saturation of such thermal instabilities
  in the zonally symmetric case.  The result is an application of
  \citet{Arnold-66} and \citet{Shepherd-88a} methods.
\end{abstract}

\pacs{02.50.Ga; 47.27.De; 92.10.Fj}

\maketitle

\section{Introduction}

A simple recipe, introduced in as early as at least the late 1960s
\citep{OBrien-Reid-67}, to incorporate thermodynamic processes
(e.g., those due to heat and freshwater exchanges through the
air--sea interface) in a one-layer ocean model consisted in allowing
buoyancy to vary with horizontal position and time, while keeping
velocity as independent of the vertical coordinate. The simplicity
of the resulting inhomogeneous-layer model with fields that do not
change with depth---referred to as IL$^0$ by \citet{Ripa-GAFD-93,
Ripa-JFM-95, Ripa-DAO-99} to reflect this---promised fundamental
understanding of ocean processes which would be difficult to gain
by analyzing direct observations or the output from an ocean general
circulation model.  A number of applications of the IL$^0$,
particularly to equatorial dynamics, appeared in the 1980s and 1990s
supporting this line of thought \citep{Schopf-Cane-83,
Anderson-McCreary-85b, McCreary-etal-97, Beier-97}.

Regrettably, the increase in computational power in the current
century has led to overemphasize reproducing observations in detriment
of gaining basic physical insight of the type that ocean modeling
based on the IL$^0$ system or variants thereof \citep{Young-94,
Ripa-JFM-95, Benilov-93, Beron-20-RMF} was expected to provide. A
pleasant surprise, however, has been to learn that layer ocean
modeling with reduced thermodynamics is regaining momentum
\citep{Zeitlin-18, Holm-etal-20, Lahaye-etal-20, Moreles-etal-21}.
Moreover, the renewed interest in this type of modeling is exceeding
a pure oceanographic interest.  Indeed, applications of the IL$^0$
have been extended to atmospheric dynamics, both terrestrial
\citep{Kurganov-etal-20} (beyond the original one \citep{Lavoie-72})
and planetary \citep{Warneford-Dellar-14, Warneford-Dellar-17}.

\begin{figure}[t!]
  \centering%
  \includegraphics[width=\columnwidth]{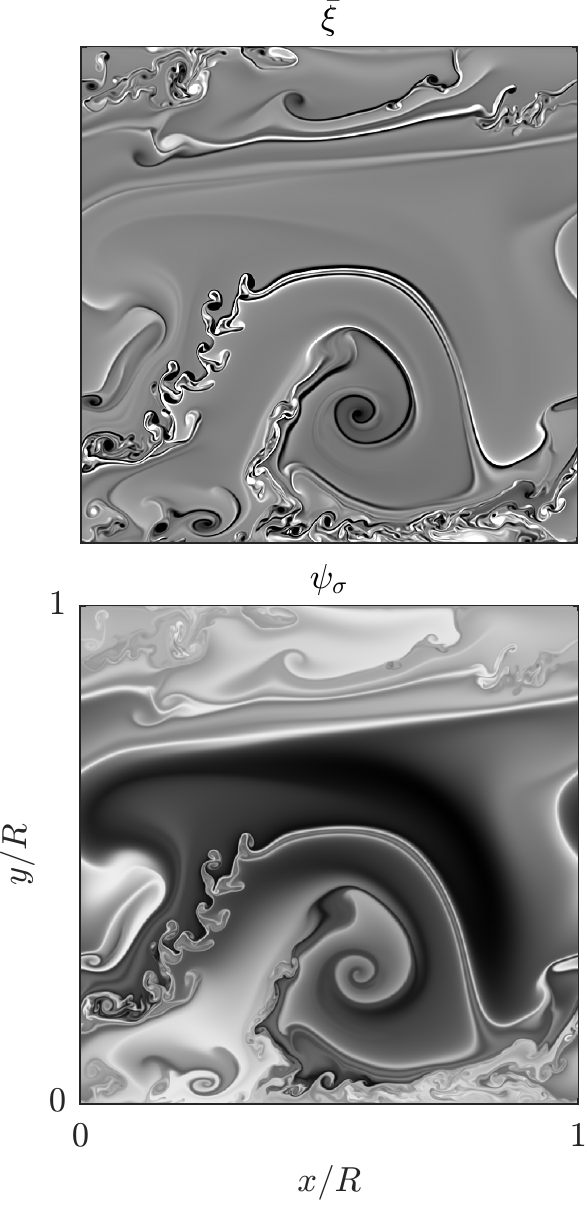}%
  \caption{Snapshot of potential vorticity (top) and buoyancy excess
  (bottom) from a numerical solution of system \eqref{eq:il0}
  initialized close to an unstable uniform zonal flow in a periodic
  channel of the $\beta$-plane using a pseudospectral code
  \citep{Beron-etal-08-JAS} on a $1024^2$-resolution grid.  Note
  the Kelvin--Helmholtz-like rollups with scales much smaller than
  the deformation radius ($R$) of the problem.}
  \label{fig:il0-rollup}%
\end{figure}

Of particular interest to the present work are the numerical
simulations of the IL$^0$ by \citet{Holm-etal-20}, which have shown
that it can sustain subinertial (i.e., with frequency smaller than
the local Coriolis parameter, twice the local Earth's rotation rate)
circulatory motions (cf.\ Fig.\ \ref{fig:il0-rollup}) that resemble
quite well submesoscale (1--10 km) features often observed in
satellite ocean color images.  This note is concerned with identifying
a mechanism that can prevent such thermal instabilities
\citep{Gouzien-etal-17} from growing indefinitely in the absence
of a high-wavenumber cutoff in the IL$^0$ model \citep{Fukamachi-etal-95,
Young-Chen-95, Ripa-JGR-96}.

\section{A quasigeostrophic IL$^0$}

Consider a low-frequency approximation to the IL$^0$ model in a
reduced-gravity setting (i.e., with the active layer floating atop
a quiescent, infinitely deep layer of constant density), which is
most appropriate to study near-surface (mixed-layer) ocean processes.
Let $x$ (resp., $y$) point eastward (resp., northward) on a zonal
$\beta$-plane \citep{Pedlosky-87} channel, $L$-periodic and of width
$W$.  The quasigeostrophic IL$^0$ takes the form \citep{Ripa-DAO-99}
\begin{subequations}
\begin{equation}
  \partial_t\bar\xi + [\bar\psi,\bar\xi] = R^{-2}[\bar\psi,\psi_\sigma],\quad
  \partial_t\psi_\sigma + [\bar\psi,\psi_\sigma] = 0,
\end{equation}
with invertibility principle 
\begin{equation}
  \nabla^2\bar\psi - R^{-2}\bar\psi =
  \bar\xi - R^{-2}\psi_\sigma -\beta y,
\end{equation}
\label{eq:il0}%
\end{subequations}
all subjected to no-flow through the coasts,
$\smash{\partial_x\bar\psi\vert_{y=0,W}} = 0$, constancy of Kelvin
circulations along them, $-\smash{\int_0^L \partial_y\bar\psi\vert_{y=0,W}}
\d{x} = \gamma_{0,W} = \const$, and $L$-periodicity in $x$. In
\eqref{eq:il0}, $[\,,]$ is the canonical Poisson bracket (Jacobian)
in $\mathbb R^2\{x,y\}$ and $R :=
\smash{\frac{\sqrt{g_\mathrm{b}H_\mathrm{r}}}{|f_0|}} > 0$ is the
Rossby deformation radius, where $g_\mathrm{b} > 0$ is the reference
(i.e., in the absence of currents) buoyancy of the active (relative
to the passive) layer whose thickness is $H_\mathrm{r} > 0$, and
$f_0$ is the Coriolis parameter at the southern coast (its dependence
on latitude is represented by $f_0+\beta y$).  The instantaneous
layer thickness, buoyancy, and velocity, $h = H_\mathrm{r}\smash{\big(1
+ \frac{\bar\psi - \psi_\sigma}{f_0R^2}\big)}$, $\vartheta =
g_\mathrm{b}\smash{\big(1 + 2 \frac{\psi_\sigma}{f_0R^2}\big)}$,
and $\u = \nabla^\perp\bar\psi$, respectively. Alternative forms
of \eqref{eq:il0} are presented in \citet{Ripa-RMF-96,
Warnerford-Dellar-13, Holm-etal-20}

The quantity $\bar\xi$ in \eqref{eq:il0} is a quasigeostrophic
approximation to the vertically averaged Ertel's potential vorticity
\citep{Ripa-JFM-95}, which is not a Lagrangian constant of the
model.  This justifies the overbar notation.  Note that by the
thermal-wind balance, which is not explicitly resolved, $\u$ would
have a vertical shear proportional to $\nabla^\perp\psi_\sigma$.
In order for $\u$ to read as above upon a vertical average over $-h
\le z \le 0$, it should be of the form $\u = \nabla^\perp\bar\psi
+ \sigma \nabla^\perp\psi_\sigma$, where $\sigma = 1 + \frac{2z}{h}$,
which justifies the $\sigma$ subscript (cf.\ \citet{Ripa-DAO-99}
for details).

System \eqref{eq:il0} preserves energy,
\begin{equation}
  \E := \tfrac{1}{2}\int|\nabla\bar\psi|^2 + R^{-2}\bar\psi^2
  \label{eq:E}
\end{equation}
(where $\int$ is a short-hand notation for integration over the
zonal-channel domain and operates on everything on its right); an
infinite family of Casimirs,
\begin{equation}
  \C :=  a_0\gamma_0 + a_W\gamma_W + \int C_1(\psi_\sigma) + \bar\xi C_2(\psi_\sigma)
  \label{eq:C}
\end{equation}
with $a_{0,W} = \const$ and $C_1,C_2(\,)$ arbitrary; and zonal momentum
\begin{equation}
  \M := \int y\bar\xi.
  \label{eq:M}
\end{equation}
Furthermore, in \citet{Ripa-RMF-96} it is shown that equations
\eqref{eq:il0} possess a generalized Hamiltonian structure
\citep{Morrison-98} on the state variables
$(\bar\xi,\bar\psi,\gamma_{0,W})$ with Hamiltonian given by
\eqref{eq:E} and Lie--Poisson bracket $\{\F,\G\} := \smash{\int
\bar \xi \big[\Fb, \Gb\big] + \psi_\sigma \big[\Fb, \Gs\big] +
\psi_\sigma \big[\Fs, \Gb\big]}$ for admissible functionals of state
$\F,\G$.  (The variational derivative of a functional $\smash{\F[\p]
= \int F(x,y,t;\p,\partial_x\p,\partial_y\p,\partial_{xy}\p,\dotsc})$
is the unique element $\smash{\Fp}$ satisfying
$\smash{\left.\frac{\d{}}{\d{\varepsilon}}\right\vert_{\varepsilon =
0} \F[\p+\varepsilon\delta\p] = \int \Fp\,\delta\p}$.) The admissibility
condition for the zonal-channel domain is
$\smash{\partial_x\Fb\vert_{y=0,W} = 0 = \partial_x\Fs\vert_{y=0,W}}$.
This bracket turns out to be the same as that for ``low-$\beta$''
reduced magnetohydrodynamics \citep{Morrison-Hazeltine-84} and
incompressible, nonhydrostatic, Boussinesq fluid dynamics on a
vertical plane \citep{Benjamin-84}; so the Casimirs in \eqref{eq:C},
satisfying $\{\F,\C\} = 0$ for all $\F$, have been known prior to
the derivation of the derivation of \eqref{eq:il0}. These integrals
play a critical role in the derivation of a-priori stability criteria,
discussed below.  The Hamiltonian formalism enables the linkage of
conservation laws with symmetries via Noether's theorem (e.g.,
\citet{Shepherd-90}).  From a more practical fluid dynamics standpoint,
it provides a framework for deriving flow-topology-preserving
stochastic versions \citep{Holm-15, Holm-etal-20} of a generalized
Hamiltonian model from its analogous Euler--Poincare variational
formulation \citep{Holm-etal-98a}, which can be used to build
parametrizations of unresolvable subgrid-scale motions
\citep{Cotter-etal-20}.

\section{Stability/instability}

\begin{figure}[t!]
  \centering%
  \includegraphics[width=\columnwidth]{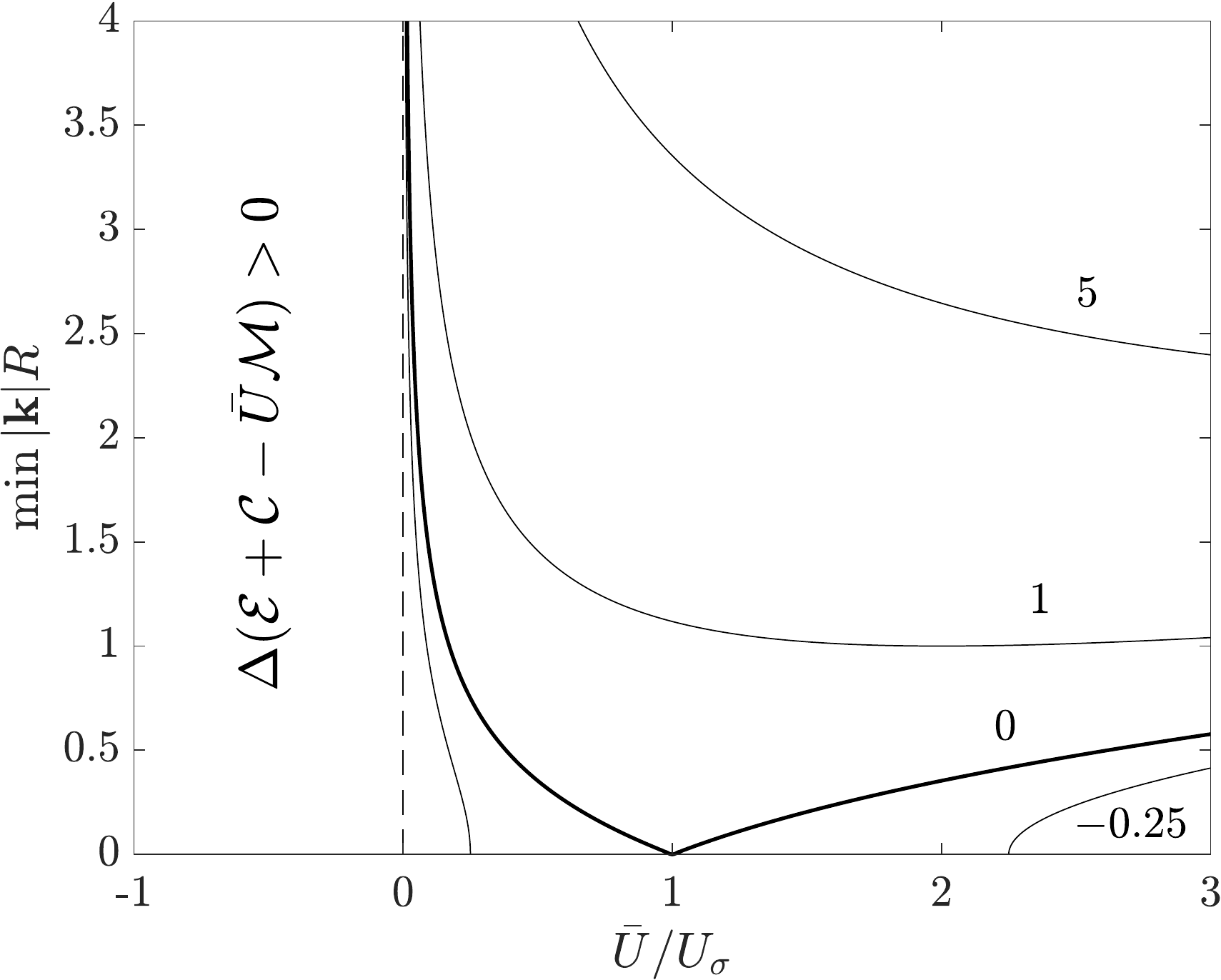}%
  \caption{Minimum wavenumber for instability of a uniform zonal
  current for various $\smash{\frac{\beta R^2}{U_\sigma}}$ values.}
  \label{fig:il0-zonal}%
\end{figure}

Let capital letters denote variables that define a basic state for
\eqref{eq:il0}, i.e., a steady solution or equilibrium to \eqref{eq:il0}
with currents. An example is the uniform zonal flow, defined by
\begin{equation}
  \bar\Psi = -\bar Uy,\quad \Psi_\sigma = - U_\sigma y,  
  \label{eq:BS}
\end{equation}
where $\bar U,U_\sigma$ are constants, with $U_\sigma <
\smash{\frac{f_0R^2}{2W}}$ so $\Theta = g_\mathrm{b}\smash{\big(1
- \frac{2U_\sigma}{f_0R^2}y\big)} > 0$. Note that, from the
thermal-wind relation, a basic state with the latter buoyancy
distribution would be consistent with a zonal flow with uniform
vertical shear given by $\smash{\frac{2U_\sigma}{H_\mathrm{r}}}$.
Thus the study of perturbations to \eqref{eq:BS} represents,
implicitly, a baroclinic instability problem of the free-boundary
type studied in \citet{Beron-Ripa-97}.  As opposed to classical
baroclinic instability \citep{Pedlosky-87}, free boundary baroclinic
instability has a soft interface, which has a slope, given by
$\smash{\frac{(U_\sigma-\bar U)H_\mathrm{r}}{f_0R^2}}$, in the basic
state. The phase speed of infinitesimal normal-mode perturbations
to \eqref{eq:BS} is given by
\begin{widetext}
\begin{equation}
  c - \bar U = - \frac{\bar U + U_\sigma + \beta R^2}{2|\k|^2R^2 + 2} \pm  
  \frac{\sqrt{(\bar U + U_\sigma + \beta R^2)^2 - 4\bar U U_\sigma(|\k|^2R^2 +
  1)}}{2|\k|^2R^2 + 2},
  \label{eq:c}
\end{equation}
\end{widetext}
which extends the result of \citet{Ripa-JFM-95, Young-Chen-95} to
the $\beta$-plane.  A sufficient condition for the absence of growing
normal modes is
\begin{equation}
  \frac{\bar U}{U_\sigma} < 0.
 \label{eq:A}
\end{equation}
In Fig.\ \ref{fig:il0-zonal} I show, as a function of $\smash{\frac{\bar
U}{U_\sigma}}$, the minimum wavenumber $|\mathbf k|$ for instability
for various $\smash{\frac{\beta R^2}{U_\sigma}}$ values. Note that
there is stability for $\smash{\frac{\bar U}{U_\sigma}} < 0$ for
all $|\k|$, as expected.  While $\beta$ can have a stabilizing
effect, the lack of a high-wavenumber cutoff of instability when
$\smash{\frac{\bar U}{U_\sigma}} > 0$ can be consequential for the
nonlinear evolution of system \eqref{eq:il0}, which however tends
to show Kelvin--Helmoltz-like circulations that saturate at
subdeformation scales rather than blowing up indefinitely
\citep{Holm-etal-20}.

Condition \eqref{eq:A} was shown in \citet{Ripa-RMF-96} to be an
a-priori condition for the formal stability of basic state
\eqref{eq:BS}, i.e., stability under small-amplitude perturbations of
arbitrary structure.  This result followed from the application of
\citet{Arnold-65} method.  This consists in constructing an integral
of motion which is quadratic to the lowest order on the deviation
from a basic state and either is positive-definite (Arnold's first
theorem) or negative-definite (Arnold's second theorem) (cf.\
\citet{Holm-etal-85, McIntyre-Shepherd-87}). In both cases, the
integral represents a norm that constrains the growth of perturbations.
For the zonally symmetric basic state \eqref{eq:BS}, such an integral
of motion is given by
\begin{widetext}
\begin{equation}
  \delta^2\H_{\bar U} := \delta^2(\E + \C - \bar U \M) = \tfrac{1}{2}\int
  |\nabla\delta\bar\psi|^2 + R^{-2}\left(\delta\bar\psi^2 - \frac{\bar
  U}{U_\sigma} \delta\psi_\sigma^2\right) 
  \label{eq:d2H}
\end{equation}
\end{widetext}
for $\C$ in \eqref{eq:C} with $C_1 = - \frac{\bar
U}{2R^2U_\sigma}\psi_\sigma^2$ and $C_2 = 0$ so $\delta\H_{\bar U}
= 0$. Note that \eqref{eq:d2H} is positive-definite when \eqref{eq:A}
holds. (That the circulation perturbations $\delta\gamma_{0,W}$ do
not enter in \eqref{eq:d2H} should not be taken as implying
positive-semidefinitness of \eqref{eq:d2H} and hence the possibility
of unarrested growth along their directions in phase space: once
initially specified, $\delta\gamma_{0,W}$ remain the same at all
times. Also note that when $R \to \infty$, i.e., the interface is
rigid, system \eqref{eq:il0} reduces to $\partial_t\bar\xi +
[\bar\psi,\bar\xi] = 0$ with $\nabla^2\bar\psi = \bar\xi - \beta
y$, for which $\bar\Psi = -\bar Uy$ clearly is stable consistent
with \eqref{eq:d2H} being positive-definite in this limit.)
Furthermore, \eqref{eq:d2H} coincides with the pseudoenergy--momentum
$\Delta\H_{\bar U}$, which is an exact integral of motion of
\eqref{eq:il0}. Indeed, $\H_\alpha$ is a Hamiltonian for the motion
as viewed from an $x$-translating frame at constant speed $\alpha$.
This shows that \eqref{eq:A} actually is a condition for the formal
stability of \eqref{eq:BS} under finite-amplitude perturbations.
However, \eqref{eq:d2H} cannot be proved to be convex, i.e., to be
bounded from below and above by multiples of an L$^2$-norm on the
perturbation field.  This precludes one from declaring \eqref{eq:BS}
stable in a Lyapunov sense \citep{Arnold-66} when \eqref{eq:A}
holds, i.e., the L$^2$-distance of a perturbation to \eqref{eq:BS}
cannot be bounded at all times by a multiple of the initial distance
(cf.\ \citet{Holm-etal-85, McIntyre-Shepherd-87}). Finally, the
possibility of proving stability when \eqref{eq:A} is violated by
seeking conditions under which \eqref{eq:d2H} is negative-definite,
namely, conditions under which $\delta^2\E =
\smash{\frac{1}{2}\int|\nabla\delta\bar\psi|^2 + R^{-2}\delta\bar\psi^2}$
can be bounded by  $-\delta^2\C = \smash{\frac{1}{2} R^{-2}\frac{\bar
U}{U_\sigma}\int\delta\psi_\sigma^2}$, is ruled out because
$\delta\bar\psi$ is not a (nonlocal) function of $\delta\psi_\sigma$
exclusively.  (On the invariant subspace of system \eqref{eq:il0},
given by $\{\psi_\sigma = \const\}$, $\delta\bar\psi = (\nabla^2 -
R^{-2})^{-1} \delta\bar\xi$.  Thus on that subspace, relative to a
general sheared zonal flow $\bar \Psi = - \int^y \bar{\mathscr
U}(y)$, both positive and negative pseudoenergy--momentum integrals
exist provided that for all $y\in [0,W]$ there are constants $\alpha$
such that $\smash{\frac{ \bar{\mathscr U}(y) - \alpha}{\beta -
\bar{\mathscr U}''(y) + R^{-2} \bar{\mathscr U}'(y)}}$ is negative
and bigger than $(\kappa^2+R^{-2})^{-1}$, respectively, where
$\kappa^2$ is the gravest eigenvalue of the Helmholtz equation with
zero Dirichlet boundary conditions at $y = 0,W$; cf.\ \citet{Ripa-JFM-92a,
Ripa-JFM-93}.)

\section{Instability saturation}

The fact that the pseudoenergy--momentum \eqref{eq:d2H} is not
convex appears to conspire against the purpose here to bound the
growth of perturbations to an unstable basic state, for which
\eqref{eq:A} is necessary violated. Let $\p$ denote the state vector.
Let the superscript S (resp., U) indicate stable (resp., unstable).
Assume that the following convexity estimate holds: $a(\PS)\smash{\|\p
- \PS\|_{t=t_0}} \le \|\p - \PS\| \le A(\PS)\smash{\|\p - \PS\|_{t=t_0}}$
for $0 < a(\PS) \le A(\PS) < \infty$ and $\|\,\|$ representing a
certain L$^2$-norm. Then one finds: $\|\p - \PU\| \le \|\p - \PS\|
+ \|\PS - \PU\| \le  A(\PS)\smash{\|\p - \PS\|_{t=t_0}} + \|\PS -
\PU\| \approx \big((A(\PS) + 1\big) \|\PS - \PU\|$ by the triangular
inequality, application of the convexity estimate, and assuming
that $\p \approx \PU$ initially, respectively. This provides a bound
on the growth of $\p - \PS$  in terms of the distance between $\PS$
and $\PU$.  Note that if there exists a convex, sign-definite
integral $\mathcal I$, namely, $0 < b(\PS)\|\p-\PS\| \le \mathcal
I[\p] - \mathcal I[\PS] \le B(\PS) \|\p-\PS\| < \infty$ for $0 <
b(\PS) \le B(\PS) < \infty$, the bound just derived follows with
$A(\PS) = \smash{\sqrt{\frac{B(\PS)}{b(\PS)}}}$.  This is
\citet{Shepherd-88a} method for finding instability saturation
bounds.  A tighter bound than the above follows by minimizing
\begin{equation}
  \mathscr B := \left(\sqrt{\frac{B(\PS)}{b(\PS)}} + 1\right) \|\PS - \PU\|
  \label{eq:b}
\end{equation}
over all possible stable states \citep[e.g.,][]{Shepherd-88b,
Shepherd-89, Olascoaga-Ripa-99, Ripa-AMS-99, Olascoaga-etal-03}.
(In a zonal channel domain bounds on the zonal-average of the
perturbation (mean flow) and deviation from it (waves) can be derived
too \citep{Shepherd-88a}, but these are not essential for purpose
of this work.)

The required convex invariant is given by the
pseudoenergy--momentum relative to a basic state with a zonal
current as in \eqref{eq:BS}, but with a more general buoyancy
distribution, viz.,
\begin{equation}
  \bar\Psi = - \bar U^\mathscr{F}y,\quad 
  \Psi_\sigma = \mathscr F(y) > -\frac{f_0R^2}{2}
  \label{eq:BS-F}
\end{equation}
provided that $\mathscr F$ has inverse, so $\bar\Psi = -\bar
U^\mathscr{F} \mathscr F^{-1}(\Psi_\sigma)$.  Clearly,
$[\bar\Psi,\Psi_\sigma] = 0$, as required for an equilibrium to
\eqref{eq:il0}.  The set of equilibria actually exceeds the zonal
flow class; however, the possibility of deriving a-priori stability
conditions using Arnold's method is restricted to this class
\citep{Ripa-RMF-96}. With a Casimir defined by $\smash{C_1 =
-R^{-2}\bar U^\mathscr{F}\int^{\psi_\sigma}\mathscr F^{-1}(\psi_\sigma)}$
and $C_2 = 0$ in \eqref{eq:C},
\begin{widetext}
\begin{equation}
  \Delta\H_{\bar U} := \int \tfrac{1}{2}\left(|\nabla\delta\bar\psi|^2
  + R^{-2}\delta\bar\psi^2\right) - R^{-2} \int^{\delta\psi_\sigma}_0
  \big(\bar\Psi(\Psi_\sigma + s) - \bar\Psi(\Psi_\sigma)\big)\d{s}
  \label{eq:DH}
\end{equation}
\end{widetext}
represents an exact pseudoenergy--momentum, which is
positive-definite when $\bar\Psi'(\Psi_\sigma) < 0$.  Furthermore
if there are constants $c_1,c_2$ such that
\begin{equation}
  0 < c_1 \le - \bar\Psi'(\Psi_\sigma) \le c_2 < \infty,
  \label{eq:L}
\end{equation}
then Taylor's reminder theorem guarantees that \eqref{eq:DH} is
bounded by multiples of the L$^2$-norm
\begin{widetext}
\begin{equation}
  \|(\delta\bar\xi,\delta\psi_\sigma)\|^2_\lambda := \tfrac{1}{2}\int
  |\nabla\delta\bar\psi|^2 + R^{-2}\big(\delta\bar\psi^2 + \lambda
  \delta\psi_\sigma^2\big),\quad c_1 \le \lambda\le c_2.  
  \label{eq:L2}
\end{equation}
\end{widetext}
This convexity estimate guarantees nonlinear stability for the basic
state \eqref{eq:BS-F} in a Lyapunov sense, i.e.,
$\smash{\|(\delta\bar\xi,\delta\psi_\sigma)\|_{\lambda,\,t > t_0}}
\le \smash{\sqrt{\frac{c_2}{c_1}}}
\smash{\|(\delta\bar\xi,\delta\psi_\sigma)\|_{\lambda,\,t = t_0}}$,
provided that \eqref{eq:L} holds, which excludes the case $\mathscr
F(y) = -U_\sigma y$ considered above.  This stability theorem enables
one to a-priori bound the finite-amplitude growth of perturbations
to any unstable basic state of system \eqref{eq:il0} using Shepherd's
method, even for the linear $\mathscr F(y)$ class, regardless of
the fact that for this specific class of equilibrium Lyapunov
stability cannot be proved. An upper bound \eqref{eq:b} will be
given by $\smash{\sqrt{\frac{c_2}{c_1} + 1}}$ times the L$^2$-distance
\eqref{eq:L2} between the basic state \eqref{eq:BS-F}, with the
condition \eqref{eq:L}, and the unstable basic state in question.

\begin{figure}[t!]
  \centering%
  \includegraphics[width=\columnwidth]{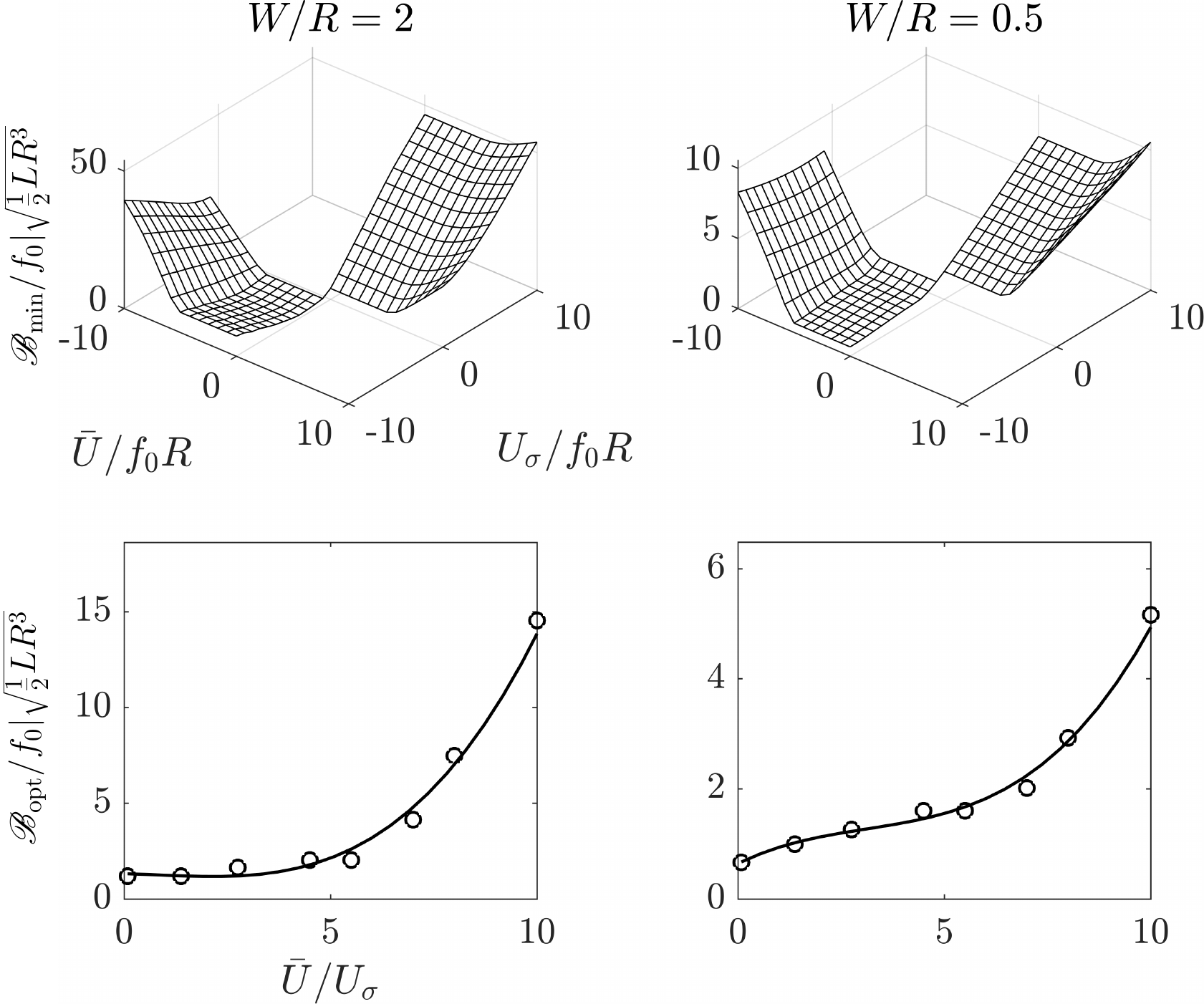}%
  \caption{(top panels) A-priori bound \eqref{eq:B} on the nonlinear
  growth of perturbations to unstable states of the class \eqref{eq:BS},
  minimized over all possible states states defined by \eqref{eq:BS-F}
  and \eqref{eq:F}. (bottom panels) Optimal bound obtained as a
  function of the basic flow's stability parameter in the unstable
  range.}
  \label{fig:bound}%
\end{figure}

As an example, consider
\begin{equation}
  \mathscr F(y) = f_0R^2\sqrt{2\exp\frac{
  U^\mathscr{F}_\sigma y}{f_0R^2}-1},\quad
  \frac{U^\mathscr{F}_\sigma}{f_0} >
  0.
\label{eq:F}
\end{equation}
This gives
\begin{equation}
  \bar\Psi(\Psi_\sigma) =
  -f_0R^2\log \frac{1}{2}\left(1 +
  \frac{\Psi_\sigma^2}{f_0^2R^4}\right),
\end{equation}
where $\Psi_\sigma$ is restricted to vary from $f_0R^2$ to
$\smash{\Psi_\sigma^{\max} = f_0R^2\sqrt{2\exp\frac{U^\mathscr{F}_\sigma
W}{f_0R^2}-1}}$ in a zonal channel of width $W$.  Its derivative
\begin{equation}
  \bar\Psi'(\Psi_\sigma) = - \frac{4\frac{\Psi_\sigma}{f_0R^2}}{1 +
  \frac{\Psi_\sigma^2}{f_0^2R^4}},
  \label{eq:A2}
\end{equation}
which is negative since $\smash{\frac{\Psi_\sigma}{f_0R^2}} =
\mathscr F(y) > 0$ by \eqref{eq:F}, making the pseudoenergy--momentum
in \eqref{eq:DH} positive-definite.  Furthermore, $-\bar\Psi'(\Psi_\sigma)$
is bounded away from zero by $c_1 = -\bar\Psi'(\Psi_\sigma^{\max})
$ and from infinity by $c_2 = 4$, implying Lyapunov stability for
the family of basic states defined by \eqref{eq:F}.  Setting
$\Phi^\mathrm{U}$ using \eqref{eq:BS} under the assumption that
condition \eqref{eq:A} is violated and $\Phi^\mathrm{S}$ using
\eqref{eq:BS-F} and \eqref{eq:F}, the bound \eqref{eq:b} on the
nonlinear growth of perturbations to $\Phi^\mathrm{U}$ with respect
to the L$^2$-norm \eqref{eq:L2} with $\lambda =
-\bar\Psi'(\Psi_\sigma^{\max})$, i.e., the smallest admissible
choice, takes the form:
\begin{widetext}
\begin{align}
  \mathscr
  B(\bar\mu,\mu_\sigma;\bar\mu^\mathscr{F},\mu^\mathscr{F}_\sigma,\nu) {}
  = & |f_0|\smash{\sqrt{\tfrac{1}{2}LR^3}} \cdot
  \left(\sqrt{\frac{\exp\mu^\mathscr{F}_\sigma\nu}{\sqrt{2\exp\mu^\mathscr{F}_\sigma\nu
  -1}}} + 1\right) \cdot \Big(\big(\bar\mu^\mathscr{F} -
  \bar\mu\big)^2\left(\nu + \tfrac{1}{3}\nu^3\right)\nonumber\\
  {} & + \left.2\nu\frac{\sqrt{2\exp\mu^\mathscr{F}_\sigma\nu -
  1}}{\exp\mu^\mathscr{F}_\sigma\nu}
  \int_0^1\left(\sqrt{2\exp\mu^\mathscr{F}_\sigma\nu \tfrac{y}{W}
  - 1} + \mu_\sigma\nu\tfrac{y}{W}\right)^2
  \d{\tfrac{y}{W}}\right)^\frac{1}{2}, 
  \label{eq:B}
\end{align}
\end{widetext}
where the unstable basic state parameters $\bar\mu := \smash{\frac{\bar
U}{f_0R}}$ and $\mu_\sigma := \smash{\frac{U_\sigma}{f_0R}}$ are
such that $0 < \smash{\frac{\bar\mu}{\mu_\sigma}} =: \tau$, the
stable basic state parameters $\bar\mu^\mathscr{F} := \smash{\frac{\bar
U^\mathscr{F}}{f_0R}}$ and $\mu^\mathscr{F}_\sigma :=
\smash{\frac{U^\mathscr{F}_\sigma}{f_0R}} > 0$, and the channel's
aspect ratio $\nu : = \smash{\frac{W}{R}} > 0$. The top panels of
\ref{fig:bound} show, estimated numerically, $\mathscr
B_\mathrm{min}(\bar\mu,\mu_\sigma;\nu) :=
\min_{\bar\mu^\mathscr{F},\mu^\mathscr{F}_\sigma}\mathscr
B(\bar\mu,\mu_\sigma;\bar\mu^\mathscr{F},\mu^\mathscr{F}_\sigma,\nu)$ for
two selected values of $\nu$.  The bound, which does not depend on
the strength of the $\beta$ effect, decreases with $\nu$.  Indeed,
it decays to zero as the width of the channel shrinks to zero, as
can be anticipated, or as $R$ tends to infinity, limit in which the
basic flow is stable as noted above.  As a function of the instability
parameter $\tau$, the bound is multivalued.  The bottom panels of
Fig.\ \ref{fig:bound} show $\mathscr B_\mathrm{opt}(\tau;\nu)$
obtained numerically by keeping, for each $\nu$, the least attainable
value per $\tau$ value (interval), which provides the tightest bound
possible for each unstable basic state (the solid curve is a
polynomial fit to the open dots).  As can be expected, the obtained
optimal bound decreases toward criticality ($\tau = 0$). Yet at
$\tau = 0$ the bound is not zero (except when $\nu\to 0$) as it
might be desired.  A choice of stable basic state different than
\eqref{eq:F} could lead to the desired result and an overall tighter
bound.

Nonlinear saturation bounds of the type above have been shown to
exist even for systems subjected to forcing and dissipation
\citep{Shepherd-88b}.  Consider, for instance, the case of system
\eqref{eq:il0} with the $\bar\xi$-equation forced (damped?) by
$R^{-2}[\bar\psi,\hat\psi]$, where $\hat\psi$ is a prescribed
function of latitude ($y$).  This system is Lie--Poisson, with
bracket as in \eqref{eq:il0} and Hamiltonian given by $\smash{\hat\H
= \H + R^{-2}\int\hat\psi\psi_\sigma}$.  (In \citet{Holm-etal-20},
$\hat\psi$ is given the interpretation of a bottom (surface in the
present case) topography; however, they do not include
the corresponding topographic-$\beta$ term in the potential vorticity
$\bar\xi$.) The system thus have the same Casimirs as the original
(unforced, inviscid) system \eqref{eq:il0}, given in \eqref{eq:C},
and the $x$-translational symmetry of $\hat\psi$ makes the zonal
momentum \eqref{eq:M} to be conserved as well.  Furthermore,
\eqref{eq:BS} and \eqref{eq:BS-F} are admissible basic states, and
all of the results above carry over mutatis mutandis to the forced
case.  (The only differences appear in the Casimir choices, with
$\smash{C_1 = -\frac{\bar U}{2R^2U_\sigma}\psi_\sigma^2 -
\frac{\hat\psi}{R^2}\psi_\sigma}$ and $C_2 = 0$ for the
pseudoenergy--momentum \eqref{eq:d2H} and $\smash{C_1 =
-\frac{\hat\psi}{R^2}\psi_\sigma - \frac{\bar
U^\mathscr{F}}{R^2}\int^{\psi_\sigma}\mathscr F^{-1}(\psi_\sigma)}$
and $C_2 = 0$ for that in \eqref{eq:DH}.)

The above provides reason to expect (hope) that the bounds discussed
here can play a role in arresting the growth of small-scale circulatory
motions developing in direct numerical simulations of the IL$^0$
system \eqref{eq:il0}, even in the forced--dissipative regime.

\section{Concluding remarks}

The result of this work adds support to thermodynamically-active-layer
ocean modeling as described by the IL$^0$ system \eqref{eq:il0},
particularly for investigating with confidence in a geometric
mechanics framework \citep{Holm-etal-20} the contribution of
unresolved submesoscale motions to transport at resolvable scales
in the upper ocean, a topic of active research \citep{McWilliams-16}.
Interestingly, the two-dimensional simulations described in
\citet{Holm-etal-20} suggest a scale separation consistent with
satellite ocean color images and statistical analysis of in-situ
Lagrangian ocean observations \citep{Beron-LaCasce-16}, but is
challenged by three-dimensional surface-quasigeostrophic simulations
\citep[e.g.,][]{Klein-etal-08}, which suggest a continuous inverse
energy cascade. The extent to which the small-scale circulations
and the associated scale separation represent a peculiarity of the
IL$^0$ model needs to be assessed, which is reserved for the future.
An appropriate framework for this is provided by models with more
vertical resolution, and hence better thermodynamics, than the
IL$^0$ model \citep{Ripa-JFM-95, Ripa-DAO-99, Beron-20-RMF}.

\section*{Supplementary material}

This paper does not include supplementary material.

\section*{Author's contributions}

This paper is authored by a single individual who entirely carried
out the work.

\begin{acknowledgments}
  The author thanks M. Josefina Olascoaga for the benefit of
  discussions on Shepherd's method.  Corrections to the manuscript
  by Daniel Karrasch are appreciated.
\end{acknowledgments}

\section*{AIP Publishing data sharing policy}

This paper does not involve the use of data.

\appendix

\section{Free waves}

For completeness, recall that the free waves of the IL$^0$ model,
i.e., infinitesimally small, normal-mode perturbations to a reference
(i.e., quiescent) state of \eqref{eq:il0} characterized by $\bar\Psi
= 0 = \Psi_\sigma$, are given by \citep{Ripa-DAO-99}: a Rossby wave,
with frequency $\smash{\omega = -\frac{k\beta}{|\k|^2 + R^{-2}}}$
and for which $\delta\bar\psi \neq 0$ $\delta\psi_\sigma \neq 0$,
and an $\omega = 0$ mode, so-called force compensating mode
\citep{Ripa-JGR-96}, with $\delta\bar\psi = 0 = \delta\psi_\sigma$
or equivalently $2g_\mathrm{b}\delta h + H_\mathrm{r}\delta \vartheta
= 0$.  Using the Casimir $\C = \smash{\frac{1}{2} R^{-2}\int
\psi_\sigma^2}$, relative to this reference state, $\Delta(\E + \C)
= \smash{\frac{1}{2}\int |\nabla\delta\bar\psi|^2 + R^{-2}
(\delta\bar\psi^2 + \delta\psi_\sigma^2) =: \E_\mathrm{f}}$ is an
exact invariant (which can called a free energy).  Being
positive-definite, it prevents the spontaneous growth of infinitesimal
perturbations to the state with no currents.  This result has
received less attention than that pertaining to the most general
reference state, characterized by $\bar\Psi = a = \const$.  Upon
choosing a Casimir of the form $\C = a\int R^{-2}\psi_\sigma +
\bar\xi$, the following turns out to be a free energy relative to
this reference state \citep{Ripa-RMF-96}: $\E_\mathrm{f} = \frac{1}{2}
\delta^2(\H + C) = \smash{\frac{1}{2}\int|\nabla\delta\bar\psi|^2
+ R^{-2}\delta\bar\psi^2 \equiv \frac{1}{2}\int |\delta\u|^2 +
f_0^2R^2 \big(\frac{\delta h}{H_\mathrm{r}} +
\frac{\delta\vartheta}{2g_\mathrm{b}}\big)^2}$.  Note that this
$\E_\mathrm{f}$ is positive-semidefinite, i.e., it can vanish for
nonzero perturbations.  More specifically, variations of $h$ and
$\vartheta$ which leave $2g_\mathrm{b}h + H_\mathrm{r}\vartheta$
(for infinitesimal perturbations this is the force-compensating
mode mentioned above) unaltered do not change $\E_\mathrm{f}$.
Spontaneous growth of such variations cannot be prevented by
$\E_\mathrm{f}$ conservation \citep{Ripa-JFM-95}.  Yet formula
\eqref{eq:B} with $\bar\mu = 0 = \mu_\sigma$ provides an upper bound
on their nonlinear growth.

\section{The axisymmetric basic state case}
 
\citet{Ochoa-etal-98} showed that unbounded $f$-plane, circular,
solid-body-rotating, lens-like, outward-buoyancy-increasing,
steady-vortex solutions to the primitive-equation set from which
system \eqref{eq:il0} derives (cold inhomogeneous so-called rodons)
are formally stable.  No steady-vortex solution to \eqref{eq:il0}
can be proved stable using the integrals of motion, given by
\eqref{eq:E}, \eqref{eq:C}, and, instead of \eqref{eq:M}, $\smash{\M
= - \int \sqrt{x^2+y^2}\bar\xi}$, which holds in the axisymmetric
case.  Indeed, only circular vortices with constant azimuthal
velocity or, equivalently, local angular velocity inversely
proportional to the radial position, which is physically odd, have
a positive-definite pseudoenergy--momentum in an axisymmetric,
bounded domain.  In the limit when $R\to\infty$ or on the invariant
subspace $\{\psi_\sigma = \const\}$ of system \eqref{eq:il0}, general
axisymmetric basic flows can be shown to be Lyapunov stable using
Arnold's method (e.g., \citet{Ripa-JFM-92b}).

%

\end{document}